# A Secure Authentication Technique in Internet of Medical Things through Machine Learning

Ahmed A. Mawgoud [a, *], Ahmed I. Karadawy [b], Benbella S. Tawfik [c]

[a] *Department of Information Technology, Faculty of Computers & Artificial Intelligence, Cairo University, Giza, Egypt,*
[b] *Radio Spectrum International Affair, National Telecommunications Regulatory Authority, Cairo, Egypt*

[c] *Information System Department, Suez Canal University, Ismailia, Egypt*

**Abstract**

The rapid growth of the Internet of Things technology (IoT) in healthcare domain led to the appearance of many security threats and risks. It became very challenging to provide full protection with the expansion in using sensor objects in medical field, this led to the Internet of Medical Things (IoMT) definition, the security part in IoMT poses a perilous problem that keeps growing; because of the data sensitivity and critical information. The lack of providing a secure environment in IoMT may lead to patients' privacy issues, not only leaving the data privacy of the patient's at risk but also their lives can be in-danger. In this paper, we providea discussion on both definition and architecture of the Internet of Medical Things (IoMT) and to propose a new authentication approach through machine learning; to enhance the security level. The authentication was developed through adopting both trust management and machine learning at the gateway for recognizing the resource-constraint devices frequencies and the access timing. The main purpose is to examine the dynamic environment of the medical field and achieve an adaptive access control. Gradually, the security improvement will be provided in IoMT systems; to reduce the communication latency, have pro-active approaches over the security risks, and provide data privacy for both the patients and doctors in a healthcare environment.

*Keywords— Machine Learning, Internet of Medical Things, IoT Security, Risk Assessment, Authentication.*

## 1. Introduction

Due to the expansion in using the IoT technology in the healthcare domain, the definition of Internet of Medical Things (IoMT) was defined as a meaning of integration between the IoT in the medical field which refers to the groups of medical devices and applications. Generally, Internet of Things (IoT) systems usually have complex architecture in which various types of devices are connected together; to provide a certain service to the end user [1]. Cyber physical systems are main parts of any IoT architecture, it integrates human interference with computer based structures and enables data-driven decision procedures [2]. Currently, IoT contains technologies such as smart grids, intelligent logistics and smart villages, improved through sensors and communication protocols. IoT provides different real-time solutions over the integration of high sensitivity sensors into medical machines. The Internet of Medical Things (IoMT) is an architecture of connected healthcare software and hardware devices, this architecture forms the medical IT systems through a network of connected nodes [3]. It can decrease needless hospital appointments and the load on healthcare systems through connecting patients to their doctors and letting the transmission of medical data over a secure network. According to Frost & Sullivan investigation, the international IoMT market worth about $20 billion in 2016; it is predicted to reach about $73 billion by 2021, at a complex annual growth rate of 26%. The market of IoMT consists of smart devices, such as wearables and medical/vital monitors, those devices can be used on the body, clinic or hospital settings; in the home or in community and associated real-time location, telehealth and other services.

Nearly 60% of international healthcare corporates have already designed and applied Internet of Things technologies, and an additional 28% are expected to fully apply their own IoT topologies by 2020 [4]. Traditional medical systems are observing a rapid change in its technology as digital transformation puts the advanced connected healthcare devices in the hands of clients and provides both patients and doctors even in remote locations better access to healthcare services [5].

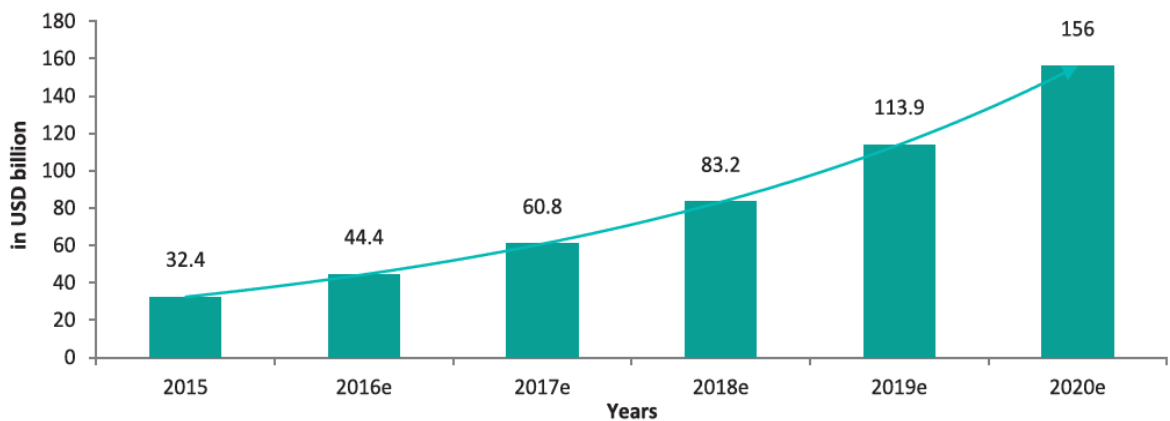

Figure 1: The investments increment rate for IoT in USB billion for medical systems from 2015 till 2020 (expected)

As shown from figure 1 above, it is expected that the market of IoMT to expand from 32.4 billion USD in 2015 to 156 billion USD% by 2020. The integrated technologies in IoMT systems are classified into three main categories:

1. Local Patient Systems.
2. End Medical Devices (i.e. Cloud Enabled Solutions, Database Systems, Device Networking and Data Management).
3. Control Systems (i.e., Firmware, Controller and Sensors).

Generally, companies such as (Software Developers, Embedded Systems, Network Operators and Semi-Conductors) are considered as the contributors for technology in IoMT ecosystem layers. Nowadays, North Americas companies are leading the market of IoMT, they provide governmental insurance regulations and IoT medical devices. In the future, it is estimated that the rise of IoT technology in medical field will be highly increased, this increment is due to the medical awareness increment (i.e. Diagnostic Facilities, Lifestyle and Disease Burden) in both American and European markets [6].

Figure 2 shows statistics about the companies' contributions in medical patents, Philips company has the highest contribution with 129 patents while both IBM and Fujifilm have the lowest contributions with only 23 and 21 respectively [7].

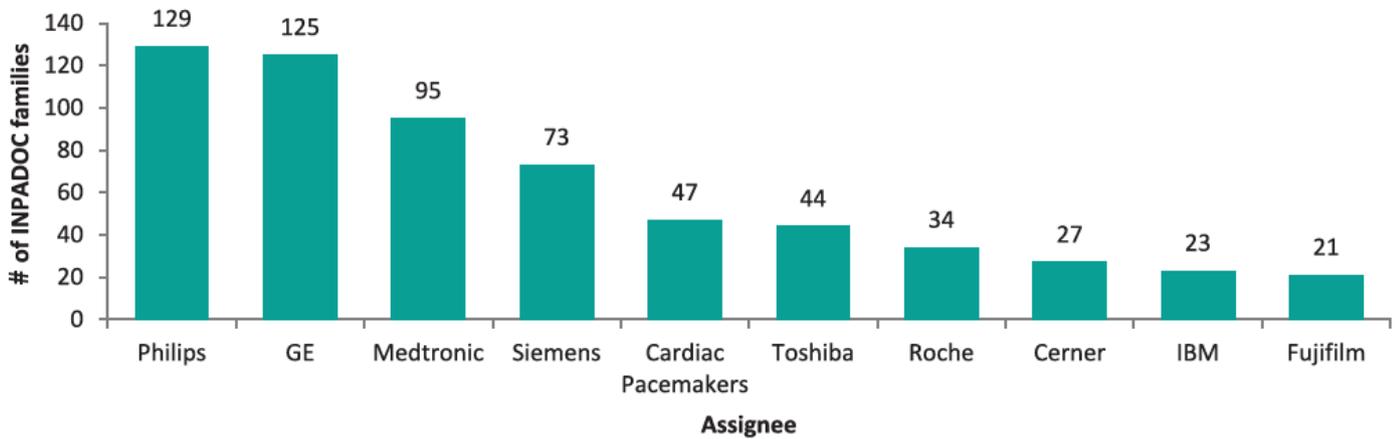

Figure 2: The number of patent that international companies have contributed the medical market with patents related to IoT solutions

Devices in IoMT are connected together in a healthcare system. IoMT technology has provide many services such as: Improving Healthcare Service, Diseases' Management, Data Analysis, Patient's Experience and Low-cost Services. According to economic research [8], the market of IoT in healthcare domain is expected to raise to about 120 billion $ by 2020. Nevertheless, the extensive diversity of IoT networks in healthcare is faced with a lot of security risks in healthcare systems. This was accredited to many reasons:

1. Medical technologies are mainly exchanging sensitive data about the patients' diseases.
2. Lack of compatibility and complex issue due to the vast number of connected smart devices in the network.
3. Privacy and security risks that face systems in healthcare environment; because of the critical data of both physicians and patients.

In emergencies, medical stuffs rush to use IoT solutions without having security part in consideration. As a result, many security problems related to confidentiality, integrity, and availability CIA appeared [9]. The rising of security risk of wireless sensor network (WSN); as all the data transmission in IoMT networks is done wirelessly. Furthermore, they security and privacy solutions of IoMT include controlling, monitoring and operating them. The breaches and vulnerabilities in the applications became an issue.

Security operations consume a huge capacity of computer resources; because of the limited resources (i.e. memory, hard capacity and power) of wireless sensor [10]. Some of these sensors have lack of encryption, the absence of encryption opens a lot of vulnerabilities and exploitations in-front of attackers in the IoMT systems. Consequently, the security and privacy issues in IoMT became a priority in the healthcare industry; because of the critical data of both patients and doctors in such a field, any kind of attack can represent a huge risk and may lead to disastrous consequences such as: life loss, indecorous treatment or financial loss. Therefore, it became mandatory to identify all the possible threats in the IoMT to support the decision making process during analyzing and designing secure solutions for IoMT environment [11]. Regardless of these efforts, IoMT still faces serious threats related to security in general as shown in figure 3, previous papers have studied the lack of privacy and security in IoT systems related to different fields in general. Nevertheless, the IoMT system differs in its security and privacy requirements because of the unique features of healthcare environment [12].

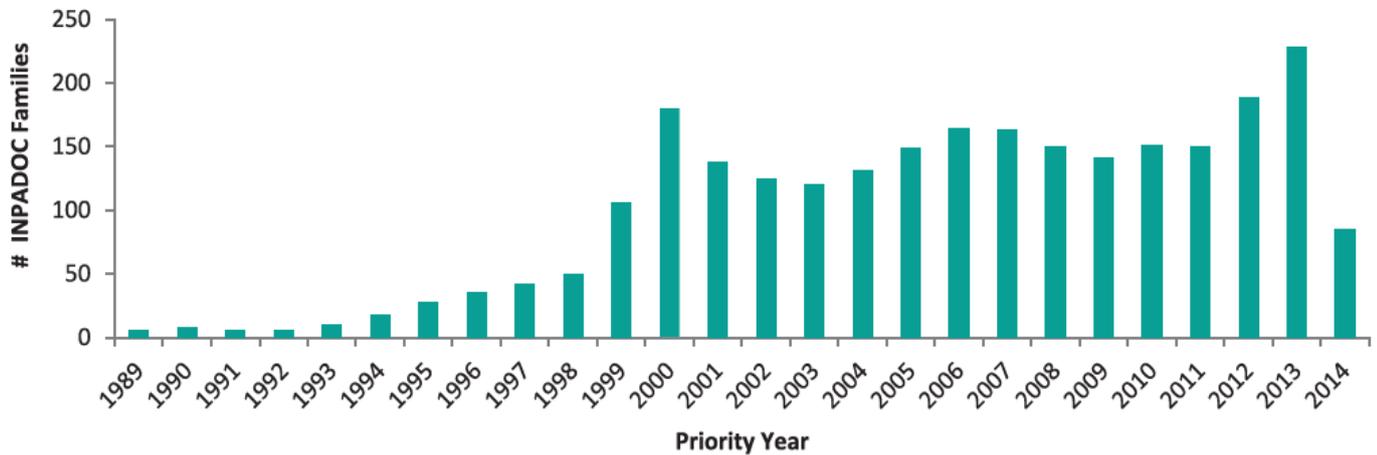

Figure 3: Statistic of the healthcare patents in medical care technology from 1989 to 2014

Moreover, the previous studies in IoT filed in general have mainly focused on TCP/IP layers. However, there was a huge shortage in investigating the risks in both physical and application layers.

In this paper, a classification about the IoT devices in smart medical systems is created; the target of this classification is to:

1- Provide an illustrated taxonomy for IoT integration into medical environment.  2-

Highlight the main risks and threats that faces IoT in healthcare field.

The threats are categorized based on (IoT layers, CIA and attack approaches). Subsequently, a proposed machine learning algorithm is tested to provide a secure mechanism for authenticated user. Finally, a conclusion which summarize the whole idea of our discussion and proposed solution along with future works [13].

## 2. Literature review

The Internet of Medical Things (IoMT) is a system consists of multiple connected nodes, each node is shaped from a set of connected IoT devices, clinical systems and wearable sensors in healthcare field. It enhances both the patient's treatment quality level and the medical time responses [14]. IoMT is part of the digital transformation in healthcare domain, the integration of IoT devices in healthcare field is driven by wireless communications, big data analytics, sensors, actuators and cloud computing. The digital transformation in medical industry has improved the targeted and personalized medicine delivery by allowing continuous communication of healthcare information [15]. Although these progressively omnipresent medical devices provide huge benefits to healthcare industry, a lot of concerns were raised about the privacy and security of these technologies. Medical systems usually collect and process critical and sensitive data about both patients' conditions to contribute in the decision making process based on this information. Cyber attackers use the existing vulnerabilities in IoMT devices in order to gain access to the medical care network and get an un-authorized contact to critical healthcare data, this is widely happening in countries in middle east and north Africa [16]. Continuous attacks on the healthcare networks can represent a serious threat on patients' life by misleading the decision making process and thus affect the treatment process negatively [17].

Figure 4 shows both advantages and disadvantages for integrating the IoT devices into healthcare systems.

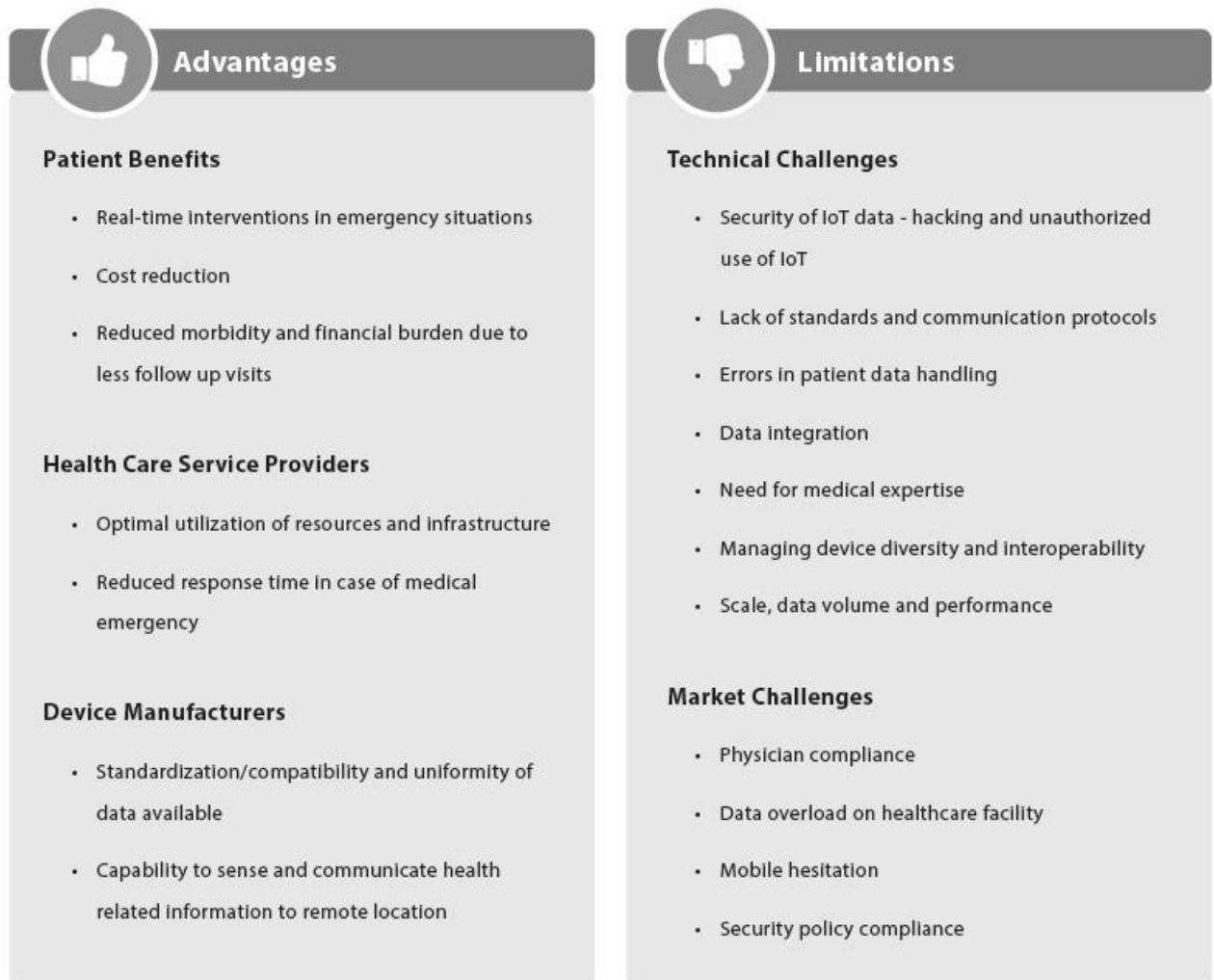

Figure 4: Both advantages and limitations that IoMT generally faces as challenges in healthcare environment

In this part, it is important to highlight state-of-the-art studies and previous researches for IoT systems in healthcare environment [18]. Contributions from medical industry and scientific researches stating innovative security and privacy solutions that were welcomed by healthcare field in general, those papers have presented trivial security solutions to cover the existing vulnerabilities and provide secure approaches for the medical devices [19].

In [20] they were studying the security-preserving SVM classification, a lot of previous approaches were proposed before, but they were only limited on securing the training set. Two approaches were presented to allow many users with various training set shards to create an SVM classifier without showing the set rate to the other users. Neither approach addresses the SVM protection after training; it is assumed that the final classifier is securely stored with a trusted third party [21 ,22].

In [23], another approach was proposed to describe the privacy support process of a trained SVM. However, this approach is only limited to Gaussian kernels; as it rejects Gaussian function roles. Although this approach provides privacy it has lack of accuracy rate; because the final SVM can cause an information leakage about the final vector classifying types. Most of the proposed methods focused on the medical classifiers, while the main goal should be releasing the final classifier during the privacy protection of the patients' support vectors. If an attacker developed a SVM classifier, they can develop the support vectors that can practically rebuild the subject's biometric [24]. Consequently, these systems are inappropriate for the discussed problem. Preferably, the main aim is to create an SVM authentication matching system which can work without revealing any data about the classified samples [25].

### 1.1 Remote Authentication Challenges

To prevent automatic authentication threats, it is difficult for the server to store huge amount of data without the user's knowledge [26]. The best solution is to provide both the client and the server with a step for every authentication over a challenge-response protocol. The main motivation is to create an authentication that provides both accuracy and privacy for the users, it is the server's responsibility to match the authenticated user independently and provide the authorization privacy for each user [27].

### 1.2 SVM Verification Protocol

In order to provide the main elements of privacy, this paper presents a developed machine learning authentication that allows a server to use the SVM classifiers for users' authentication in a secure method. The SVM classification approach conserves the users' privacy; as every classifier can have its own encryption through the user's key [28]. To avoid affecting trust on the user's authentication decisions, the server calls the classifiers to create the user's test, then the user should demonstrate to the server that it truly has the feature which is being classified. Consequently, only the user has the ability whether to decipher or to use the SVM classifiers as well as testing the feature vector [29]. The server's part is to produce a symbol that the only the user can originate from the appropriate SVM classifiers. An approach for token generation is to force the user to differentiate between the real and the non-real SVM classifier. The user's decision becomes the authentication response; because the attacker will not be able automatically to differentiate between the real and non-real SVMs as it cannot make the right response unless it was randomly made [30].

## 2 Internet of Medical Things Taxonomy

Internet of Medical Things Taxonomy is presented in figure 5, the presented classification is based on many different measurements, IoT layer, CIA, attack level, threat impact and penetration origin. The classification is updateable and there is a room for expansion by the passing time based on both new devices and attacks [31].

Figure 5 illustrates the taxonomy of IoMT systems adapting to the healthcare environment.

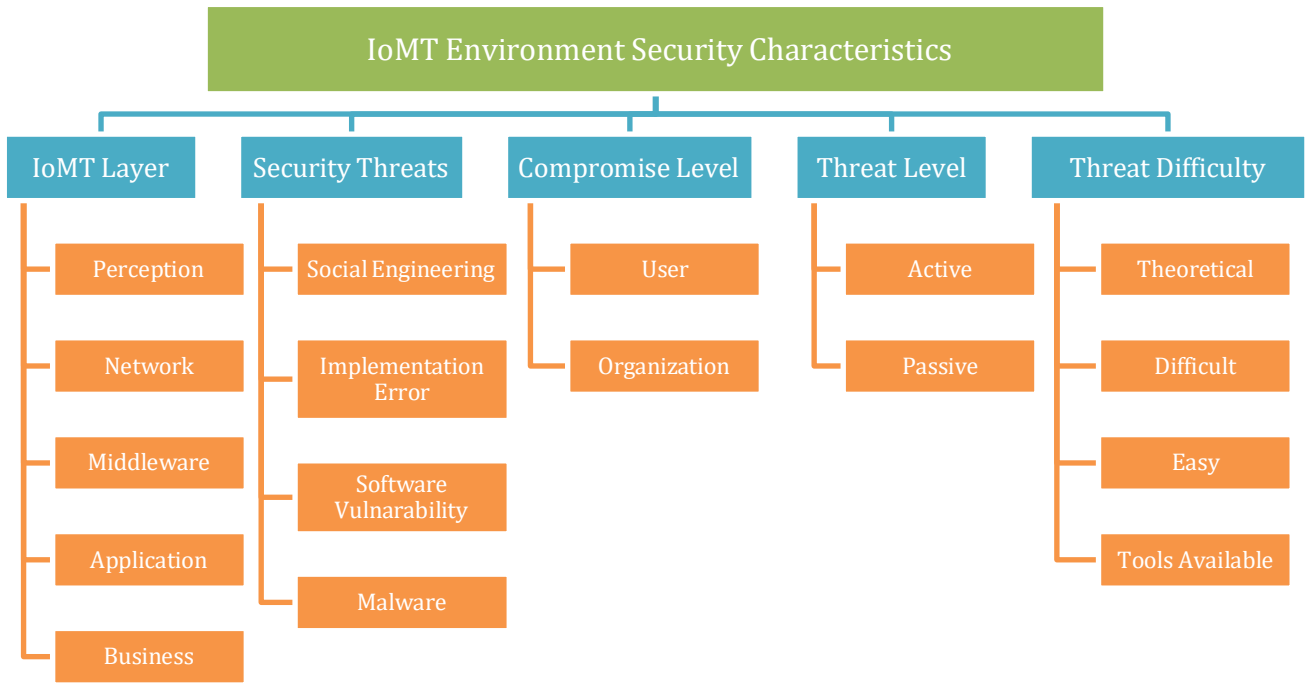

Figure 5: A hierarchy that describes the IoMT layer along with different characteristics of security threats

## 2.1 IoMT Layers

Because of the criticality of healthcare field, it became vital to manage a huge number of IoMT devices connected over the internet in heterogeneous to provide reliability. Consequently, the need has raised for a flexible design architecture. The main five IoT layers as presented in [32] provides a description for each layer with illustrating its main functionality. For every function in IoT layer there is its own security threat [33]. Hence, a representation of the security and privacy issues based on its occurrence rate in each layer:

### 2.1.1 Perception Layer

This layer has main responsibility in data collection (i.e. heart rate, pressure, temperature, etc.) using sensors in the physical layer, after that, the collected data is being transferred to the network layer. As an example, in the patient care systems, a lot of sensors are connected to the patient's body to make sure that there is a monitoring to the patient's condition and provide help when its needed [34].

As shown in figure 6, a portrait for the user interaction with the medical devices, this is done starting from collecting data through transmitting them ending with issuing the final report.

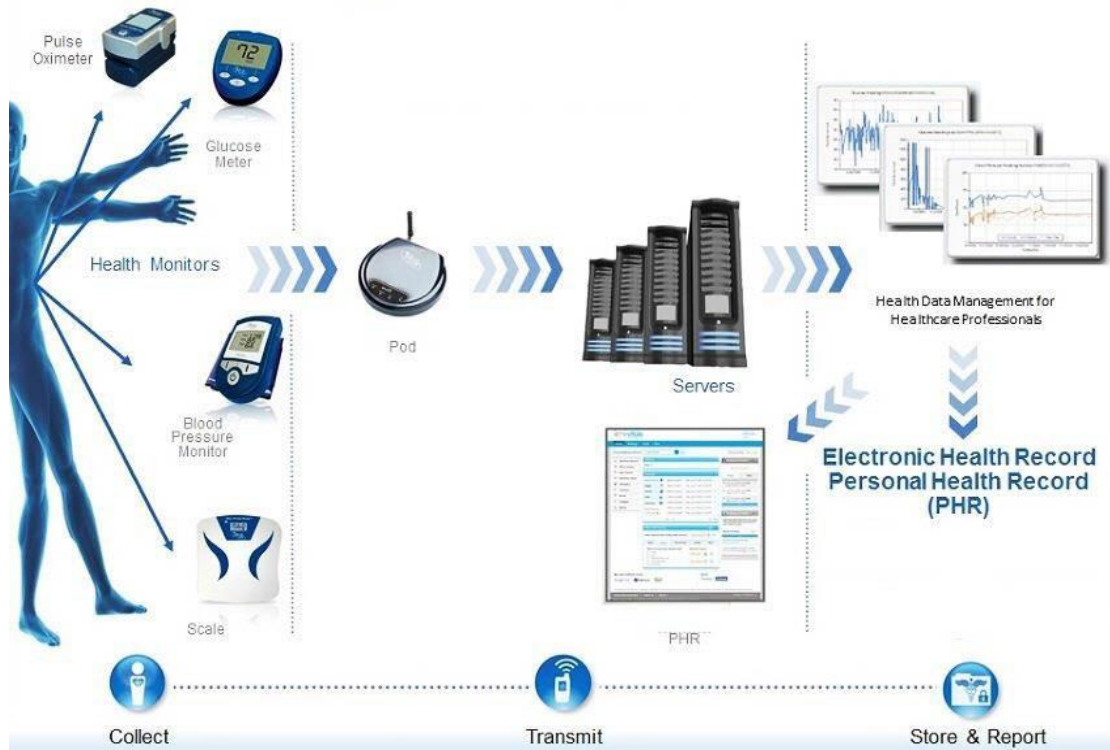

Figure 6: User interaction with software and hardware exists in the Internet of Medical Things (IoMT) environment

Table 1 represents the perception layer devices in internet of medical things.

TABLE 1: Description of perception layer devices in Internet of Medical Things

| IoMT Device | Function |
|---|---|
| Locator | Tracing the patients' location. |
| Temperatures Measureme | Providing results about body temperatures. |
| Blood Pressure | Monitoring the patients' pressure. |
| Biometric Sensor | Uniquely identify the patients' identity |
| Heart Monitoring | Using both electrocardiography (ECG)) for monitoring the heartbeats. |
| Respiratory Rate | Monitoring the breathing rate. |
| Activity Monitoring | Using gyroscope sensors for detecting patients' activities like (eating, sitting or sleeping). |
| Electronic Cardiogram | Assess the main functions of the heart for ensuring safety. |
| Pulse Oximeters | Measuring both pulse and oxygen rate. |
| Biochemical Sensors | Detecting biochemistry and harmful mixtures in the air. |

Those devices will be classified in four categories as following:

- **Implantable Devices*:* These types of objects intents to be inside the body of the patients for medical purposes, devices such as (Embedded Cardiac and Swallowable Camera Capsule) [35].

- **Tampering Devices:** Sensors are hardware that can be manipulated or tampered physically by attackers, they can affect their functionality whether by stopping those sensors or modifying their configurations. Any medical equipment with USB port has a risk to be damaged if the attacker whether through plugging in an external device to destroy its functionality or by using vulnerabilities to control the equipment through installing a malware [36].

- **Sensor Tracking:** Health monitoring equipment have GPS sensors for sharing patient's location live in emergency cases. If the equipment is not secured, attackers may spoof the data of GPS for tracking the patients' location [37]. Similarly, any sensor that uses fall detection or remote monitoring has its exploitations for revealing the patients' sensitive data [38].

- **Stationary Devices:** This class contains smart objects that are immobile and do not accompany the user all the time. Types of devices like:
  a) Surgical Objects: Used tools by medical staff while performing surgeries and origins transplants [39].
  b) Imagining Objects: Creating a visual simulation of the patient's interior body (i.e. MRI and X-rays) for medical analysis studies [40].
  c) Wearable Devices: Wearable objects that are being used by both the patient and the doctors to enable continuous monitoring and providing accurate results [41].

- **Ambient Devices**: These objects aim to gather data about the patients' surrounding area for monitoring activity forms such as breathing, sleeping, eating, walking and providing warnings to medical stuff when suspicious activities are detected, because of these types of sensors it is expected to participate in making the medical standards safer and smarter in healthcare industry, below are some examples for ambient sensors:
  1. **Pressure Sensors:** Managing both fluid and air rate in the room.
  2. **Temperatures Measurement**: Providing results about body temperatures.
  3. **Doors Sensors:** Sensing the door condition (open-close) for helping conditions with Alzheimer.
  4. **Motion Sensors:** Movement detections in small rooms.
  5. **Daylight Sensors:** Adjusting the lighting power inside the room to suit automatically the natural light.
  6. **Vibration Sensors:** Analyzing the body activities during staying in bed.

### 2.1.2 Network Layer

This layer has a responsibility for routing packets from senders to destinations and network addressing. The devices used in this layer for IoMT are as following:
- *Wi-Fi:* IoMT uses Wi-Fi systems for connecting the gateway to the end-user. IoMT devices can be stationary due to their continuous need for reliable power source [42].
- *Radio Communication:* Many IoT devices with low power -during their connection with end-users and other nodes- are using radio spectrums like 3G, LTE and Bluetooth. As an example, the IoT devices in hospitals and clinics connects other devices with each other through Wi-Fi or low powered wireless personal area network (6LoWPAN) [43].

### 2.2.3 Middleware Layer

This layer is responsible for gathering and classifying the data coming from the devices in perception layer achieving service discovery and controlling the devices access. Cloud technology became commonly used in IoMT environments (Hospitals, Clinics, Healthcare

Centers and Sanatoriums) [44].

Figure 7 below shows the relation between the physician and the patient with their existence in IoMT environment.

- **Cross-Site Request Forgery:** It is considered the most common attack in RESTful- based IoT architecture, the CSRF manipulate the user's application, this is done by making modifications by using a vulnerability in the web interface, without proper configurations the web interface in IoT layer can becomes defenseless to the CSRF attacks [45].
- **Session Hijacking:** This type of attack is common in RESTful based IoT networks; this is due to the mechanism that some IoT devices use to initiate the session connection to the layer of web interface [46].
- **Cross-site Scripting (XSS):** The XSS can inject side scripts to evade access controls through the IoT web devices web pages; this is mainly done to exploit RESTful IoT systems [47].

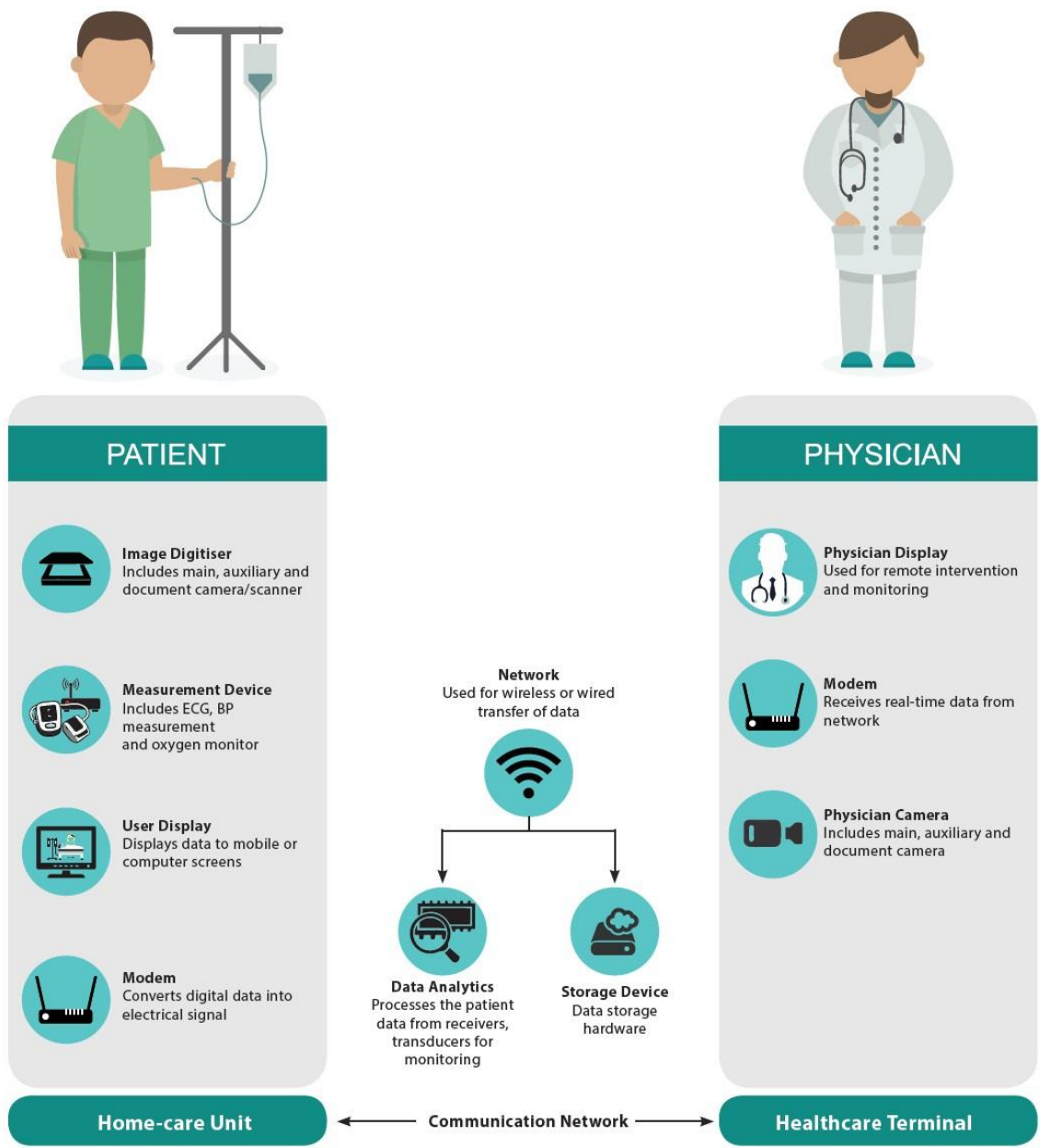

Figure 7: The patient and physician interaction through communication network using IoT devices

### 2.2.4 Application Layer

This layer represents the interface that the users interact with the surrounded IoT devices over the Middleware layer for managing, controlling and interacting [48]. Due to the large amount of data, healthcare industry had a trend on moving to the cloud platforms for scalability, integrity and cost affordance [49]. Consequently, the potential attacks' risks increased to have negative influences in the healthcare field through different types of sophisticated attacks [50]. The potential threats on this layer can be presented as follow:

- **SQL injection:** SQL injection threat occurs when the attacker attempts to use a vulnerability in the backend of the software database, this is done through inserting a malformed SQL statement. This attack represents a major threat to the IoT devices, particularly in the medical sector, as an operative SQL injection has the capability to as an effective SQL injection attack has the steal patients' information or modify sensitive data, which may cause a severe consequence in the healthcare environment overall [51].
- **Account Hijacking:** Various IoT devices interact whether with non-encrypted or weak encrypted channels at the network layer, the attacker has the ability to execute account hijacking through interrupting the packets during the authentication process of the end user. Old operating systems suffers from primitive vulnerabilities which are the main aspect for such an attack [52].
- **Ransomware:** Ransomware encrypts critical information and asks for an enormous payment for the data decryption. This risk can begin with one device then it spreads through the rest of the network. Attackers can encrypt critical information such as patient's information and physicians' data, this is mainly done to exchange the decryption key for money [53].
- **Brute Force:** This approach depends on guessing passwords through trying all the possible characters, alphabets and numbers. The IoMT applications are vulnerable to brute-force attacks. brute-force attacks as limited protection exists to prevent such threats in IoMT devices. This is accredited to the sensors' derivative computation energy [54].

### 2.2.5 Business Layer

The main responsibility of this layer is to handle the business logic for the medical provider as well as to support the business lifecycle (i.e. managing, observing and adjusting) business procedures. It is also having a responsibility in gathering information from the IoMT environment as shown in figure 8 below. The impact of cyber-attacks on this layer is very high due to the medical data sensitivity [55]. Those attacks can be the main cause for information disclosure and deception; as information disclosure or deception [56].
- **Information Disclosure:** Illegal access to credential data can cause a huge violation in the confidentiality of the IoMT network, the attacker can use any vulnerability to gain access to create, modify or even delete data related to the patient's system in the healthcare domain which can cause a severe damage [57].
- **Deception:** Infected data has the ability of disturbing data integrity and can lead to disastrous penalties. Threats such as Sinkhole and man in the middle can lead to data deception. About 56% of medical institutions do not provide a proactive approach to face such a threat.

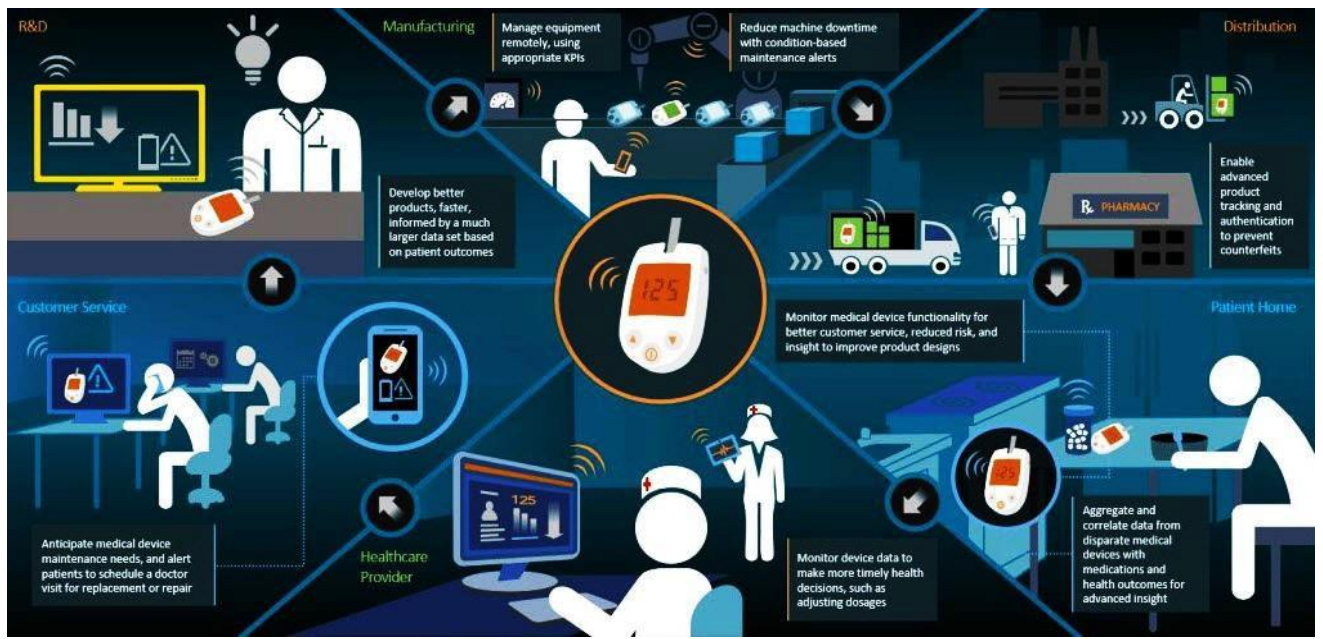

Figure 8: A representation of the functionalities of IoMT systems through different departments

## 3    Security Risks in IoMT Environment

The attack classification in the IoMT environment is relying on its damaging level after the attack is successfully done. As a result, the consequences of various cyber-attacks in medical field demonstrates a strong need to come out with proactive approaches for securing the IoMT environment. The cyber-attack threat in healthcare industry may lead to:

- **Life Risk:** Any attack occurs on devices connected directly to the patient may affect its functionality and the patient put the patient's life in-danger [58].
- **Data Exposure:** When an attacker uses an exploitation in a medical device or software, it may lead to patient's data exposure, which violates the common roles of data privacy in medical field [59].
- **Reputation Loss:** The occurred damages because of an attack whether in device or software may lead to brand value loss because of losing integrity and violates common medical privacy rules [60].
- **Financial Loss:** The resultant damages by attacks on IoMT networks needs to make damage control followed by a recovery plan. This kind of processes lead to extra budget that affect the financial part negatively in the healthcare organization [61].

TABLE 2: Description of Attacks Threats for Each Layer in IoMT Environment

| Layer | Attack type | Risk level | Description |
|---|---|---|---|
| **PERCEPTION LAYER** | Side Channel | High | An exploitation that is used for collecting data about devices actions while performing a cryptographic operation and use the collected data for reverse engineer the device cryptographic function. |
| | Sensor Tracking | High | This attack occurs when the user access a webpage, it send queries to the device sensors through automated processes that run after the page is loaded. |
| | Tag Cloning | Low | It is an easy for the attacker to capture one node within the whole network and create a new re-programed cloned node from the extracted information from the captured node. |
| **NETWORK LAYER** | Sinkhole | High | It is a type of attack in which a node tries to entice network traffic through manipulating the default routing with a fake one. One of its impact is that it can be an open door for other attacks (i.e. Selective Forwarding Attack, Spoofing and Altering Routing Information). |
| | Rogue Access | Low | It is a wireless access point that is installed on a secure system without obvious permission from a local network admin, it can be added by an internal authorized employee or by a malicious attacker. |
| | Denial of Service | High | It is a cyber-threat in which the attacker looks for to cut the availability in network resources to the targeted users through temporarily or permanently disrupting the host services. |
| | Man-in-the-middle | High | It is an attack in which the attacker surreptitiously transmits and can possibly modifies the connections between two users who are directly connecting with each other. |
| **MIDDLEWARE LAYER** | XSS | Medium | Cross-Site Scripting (XSS) cyber threats are a sort of injection that malicious scripts are being injected by the attacker in a trusted website. |
| | Session Hijacking | High | It happens when the attacker compile commands through an active connection between two nodes using source-routed IP packets, the command output hiding itself as a one of the normal users. |
| | CSRF | High | It is a type of attack which forces an end user to auto execute scripts on an authenticated web software. |
| **APPLICATION LAYER** | SQL Injection | Medium | The attackers tend to use a SQL injection into the database server through using vulnerabilities in order to gain access to the whole system. |
| | Brute Force | High | This technique mainly occurs through guessing the password characters and numbers through using a key derivation function. |
| | Account Hijacking | High | A sort of identity theft, in which can be used to gain and the financial and social consequences can be shattering. |
| | Ransomware | High | It is a malware form that rogue software code efficiently holds a user's PC hostage till a ransom financial request is paid. Ransomware often penetrates a PC as a system Trojan which can take an advantage of high security vulnerabilities. |
| **BUSINESS LAYER** | Disruption | High | It is a risky threat arising from direct or undirected incidents which is a main cause for a security breach, IoT devices damaging and networks. |
| | Information Disclosure | High | It occurs when a software fails to provide protection for the critical and confidential data from the access to a certain information in the ordinary circumstances. |

## 4 Problem Statement

In a traditional authentication, if the server finds a match between the user's inputs with the stored data, it is the server's obligation to keep the privacy and the security of this authentication operation. Additionally, if an eavesdropper catches the user authentication data that is being stored on the server.

As shown in figure 9, the possible attacks and security challenges that can be a result of disastrous concerns and cause a deep damage in the network of IoT [62]. This is mainly happening due to precarious parts the IoT presents in providing support for different applications through the connection of immense heterogeneous devices and industry operations, in addition to sequential response from the huge symmetric correlation in IoT. Furthermore, the usage of (Resource-Constrained Devices) can be risky and represent a huge threat to the whole IoT network through forging, tampering, data injection, and spoofing attacks, with the consecutive effect, those risks can lead to a IoT system failure [63]. Specifically, for the applications that depend in cooperation with different entities.

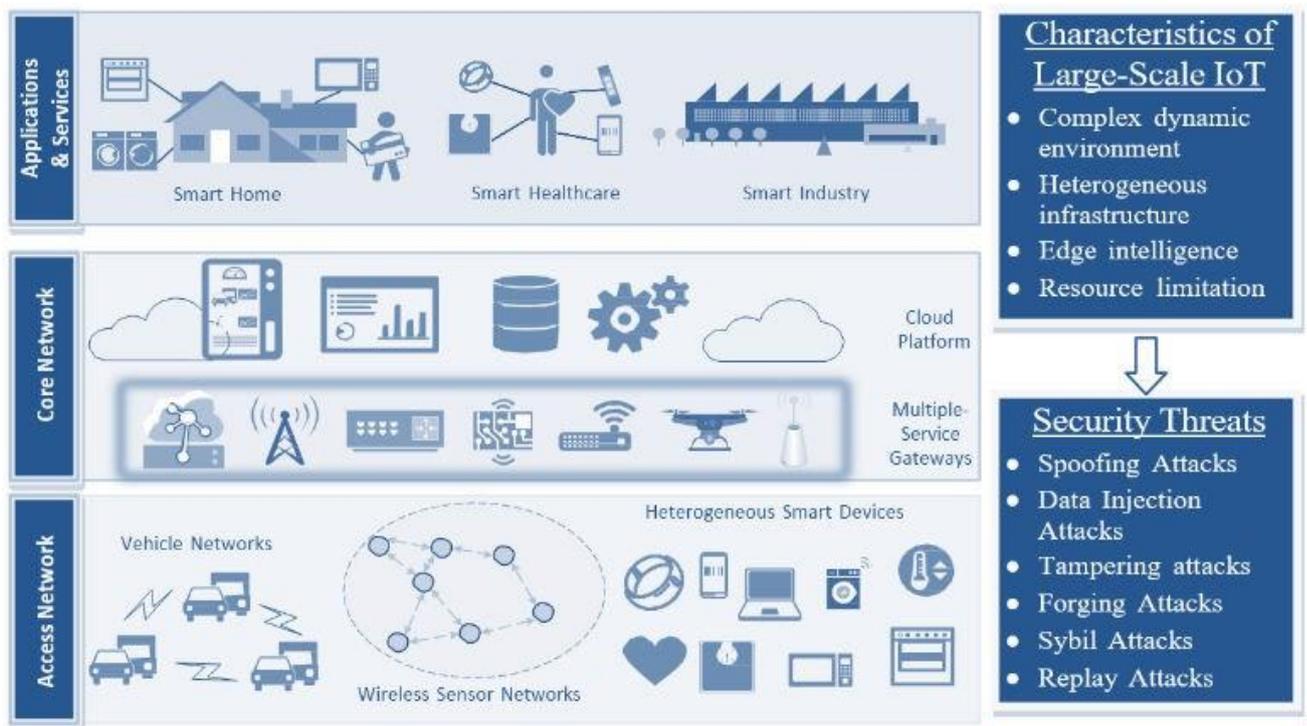

Figure 9: Characteristics & Security Risks in Internet of Things

Authentication has been defined as key security mechanism for IoT architecture design; as the hackers need to have an access to the IoT system to start their attack. This technique secures the communication within the IoT network through approving the identities and their right for accessing to the authorized resources in the network. Nevertheless, the unique features of IoT networks bring vulnerabilities in security provisioning [64]. Obviously, IoT devices that suffer from resource restrictions would not provide the required security mechanisms for the marvelous devices included in the IoMT systems demand low-delay transmissions to assure the performance of their communication. Therefore, to provide protection against those security challenges, this paper focuses on examining the challenges of the traditional authentication schemes in large IoT systems and providing approach for enriching security in IoT systems [65]. The traditional authentication methodology including physical layer key generation mechanisms and cryptography mechanisms, can suffer from both high latency and complexity and can fail in adapting in complex dynamic environment specially in IoMT systems. Significantly, a leakage may occur to the generated keys during the security management operations. The generated keys challenges in the IoMT systems [66]. The traditional cryptography methods need protracted process and improved overhead; in order to increase the security level. Therefore, it will lead to high overhead in both computation and communication. There are unendurable for IoMT system that have expressively high number of smart devices and resource-constrained machines with synchronized communications. Furthermore, the traditional statistical techniques for authentication need sufficient time and high resources for attaining the statistical resources. Consequently, it will lead to a limited proficiency in detecting attacks instantly [67]. Thus, there is a necessity for having a lightweight authentication approach to be provided for the applications in the IoMT systems. Concisely, failing to adapt a dynamic environment can be because of lack of security; so enriching the security is highly needed in IoMT networks, particularly when it comes to the data that uses artificial intelligence.

## 5 Proposed Solution

The proposed technique makes use of machine learning-enabled models in IoMT networks; as a method for providing both security, management and achieving secure authentication, through providing an authentication model [68]. In this technique, machine learning participates in securing IoMT systems in general, not just by gathering and using important data but also providing authentication self-adaptation [69].

*5.1 Support Vector Machine (SVM) in authentication*

To identify multiple IoMT sensors objects based on their pseudo-random accesses in both domains of frequency and time. In case, when the value of the time domain or frequency domain is identical to its distinctive Pseudo-Random Binary Sequence (PRBS), then the gateway will authenticate the device.

The kernel for PRBS generation between the sensors and gateways can be gained by using its distinctive attributes, as demonstrated in physical layer attributes in [70], along with discovering the Support Vector Machine (SVM) to classify the attribute measurements in non- linear approach. Therefore, this approach provides fast authentication through recognizing the time or frequencies and security for authentic communications without the need for complicated computations to establish an authentication approach [71]. Mainly, the selected attributes measurements are obtained through channel penetrating, to convert the measurements to binary structures, a quantization algorithm was developed based in the SVM to develop an optimum non-linear approach to separate both sparse and dense data.

In comparison with other approaches based on Received Signal Strength (RSS) such as the one presented in [72], the proposed approach in this paper decreases the wrong decisions rate by decreasing the boundary measurements. After that, the gateway transmits the non-linear boundaries to the sensors. As a result, the identical sequel will be attained on both sides of the sensor and gateway due to the channel exchange.

Verifications are made using hash functions, to obtain the matching seed and then the identical PRBS will be generated to authenticate both gateway and sensor [73]. Obviously, the generated seed is hidden from any other object due to the unpredictable and unique nature of the used communication link. In this approach, the machine learning technique (i.e. SVM) simplifies the security improvement by obtaining a non-linear classifier at the gateway that provides the needed storage and energy for training. Decisively, because of the consequential optimal non-linear confines by SVM, the extreme identical binary structures are needed in the quantization stage. Thus, there is no need for seed transmission for verification [74]. Consequently, in the proposed authentication IoMT network approach is at gateways.

A proposed authentication technique is provided as illustrated using a workflow as illustrated in Figure 10 below.

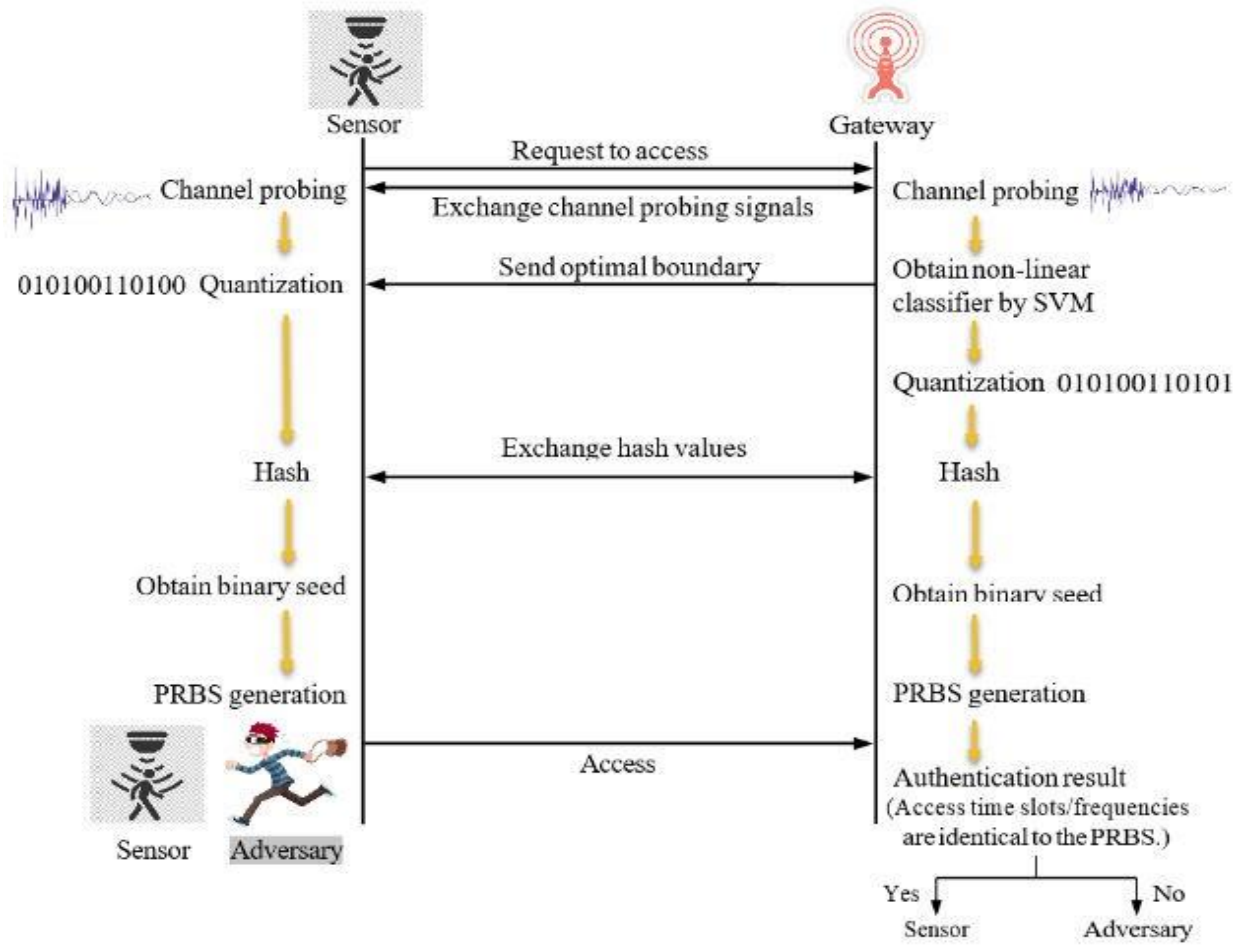

Figure 10: A diagram illustrates a Developed Authentication Technique using SVM

## 5.2 Technique based on Machine Learning and Trust Management

To achieve secure authentication, we visualize a security management access based control technique by perceptively exploiting the transmitter time change attributes (i.e. users' behaviors, communication attributes and hardware attributes) for improving security [75]. Specifically, one attribute is used for primary authentication. Therefore, it will achieve transmitter fast access, to measure the security threats caused by non-detected antagonist, we discover the trust management to launch the protection relation between transmitter and receiver through dividing the resources of IoMT into X levels [76]. The transmitter is synchronized to access the authentication layer conformable to its trust value and subsequently in an approach [77]. The machine learning was designed to be updated automatically based on the estimated attribute real-time detection at the receiver. To achieve this, the decision-making in a high layer is essential for demonstrating the authentication to enhance security performance [78]. Through this approach, the uncertainty threats caused by attackers can be controlled in real-time to achieve progressive security for IoMT devices. The approach based on machine learning in a dynamic IoMT environment is proposed in the following algorithms.

---

**Algorithm:** Machine Learning depend on Authentication

---

**Input:** Attributes Measurements
**Output:** Authorization Level
1: // Primary authentication:
2: Selecting attribute for initial authentication, e.g. IP Address, Channel, Hardwar Biometric Attribute;
3: Get the primary trust rate of the transmitter to check its validation to authenticate;
4: // Progressive authentication:
5: Select many attributes for authentication;
6: **While** the Training Data is available,
7: **Do**
8:     **If** new input is identified as an outlier
9:     Trust Value $q[x]$ decrement;
10:     **Else**
11:     Trust Value $q[x]$ increment;
12:     **End If**
13:     **If** Trust Value is part of the set
14:       Approve this Transmitter with $y$-level of resources;
15:     **Else**
16:       Terminate the Transmitter connection;
17:     **End if**
18: **End While**

---

Furthermore, the benefit from collaboration between devices can provide smooth protection and enrich security in IoMT networks. As an example, exploitations can happen for IoMT devices while updating the trust relations between the transceiver [79]. As a result, there is a need to achieve a high comprehensive assessment for the transmitter. An exploration using game theory can be done to study the interactions nature between the devices to attain co-operation between devices as well as managing trust adaptively [80].

The main idea is to have different treatment levels for misbehaved devices. By pursuing the equilibrium, devices in the IoMT network should be having a moderate behavior pattern [81]. Authentic devices can have the highest level of authorization in the network even with the adversaries' presence; in order to guarantee a secure communication in IoMT networks. In this method, the contribution of the machine learning methodologies for security part is by keep learning about different transmitter behaviors and cooperating with IoMT object under the limitation of statistical properties [82].

# 6 Experiment

Table 1 represents the results of the comparison between physical layer key generation approach and authentication approach. It also shows the static nature of the physical layer key generation approach while the authentication approach has a dynamic nature that can achieve constant authentication in the time-domain. Furthermore, there is no need for seed transmission to be part of our method to provide the highest possible rate of protection for the generated PRBSs. A potential key leakage may occur because of the key transmission in the physical layer (Key Generation) model during the process of security enhancement. However, there is a noticeable low rate of communication achieved by the developed authentication model during the key generation process; as it needs time due to the complex computation process to generate PRBS at the gateway.

TABLE 2: A COMPARISON BETWEEN PHYSICAL LAYER KEY APPROACH AND AUTHENTICATION APPROACH

| Type | Key Generation Approach | Authentication Approach |
|---|---|---|
| Characteristic | Static | Dynamic |
| Key transmission | Valid | Non-valid |
| Channel Penetrating | Valid | Valid |
| Quantization | Valid | Valid |
| Data verification | Valid | Valid |
| Privacy Augmentation | Valid | Non-valid |

However, in the proposed approach for x authentication times, the time complexity is lower than the one that was presented in [83], which lead to the reason of the importance of fast authentication in enhancing security in IoMT systems. Figure 11 below shows the observation of many different attributes. The X-axis represents the number of the adversary attacking behavior obtained by the receiver while the Y-axis represents the trust value. The trust value update for the authenticated devices (with different attributes):
- **RSSI:** Received Signal Strength Indication.
- **CFO:** Carrier Frequency Offset.
- **Attacking Behavior.**

In the proposed authentication method, there are three levels of authentications:
- **Level 1:** A set that includes null resources\services.
- **Level 2:** A set that includes limited resources\services.
- **Level 3:** A set that includes all resources\services.

From figure 11 below, it is noticeable that the rate of attacking behavior is lower than the rate of received signal strength indication or the RSSI & the carrier frequency offset.

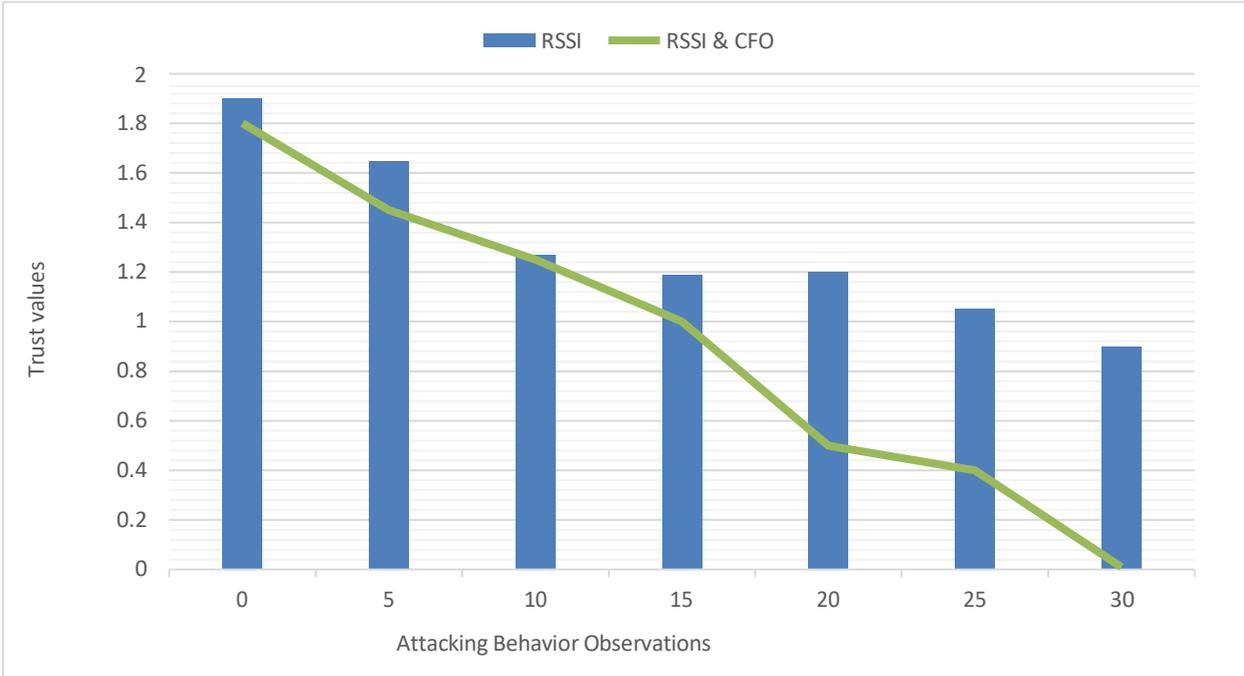

Figure 11: Trust Value Update for mis-detected adversaries for RSSI & CFO

Precisely, malicious attacks represent the attacking behaviors; those behaviors can be controlled by switching them on or off, this switch will be based on applied communication nature in the IoMT systems. The creation of preliminary trust value is based on receiver's IP address evaluation. There is a possibility for the sender or the receiver to fake the IP address, from figure 12 it is detected that there is a noticeable fast decrement of the trust value for detected adversary by adding more attributes for the holistic authentication, as a result, the average value was set to be with a value of 0.9. The main reason for this tendency is that the sender has a high confidence level of both the attacking behavior and the authentication attributes.

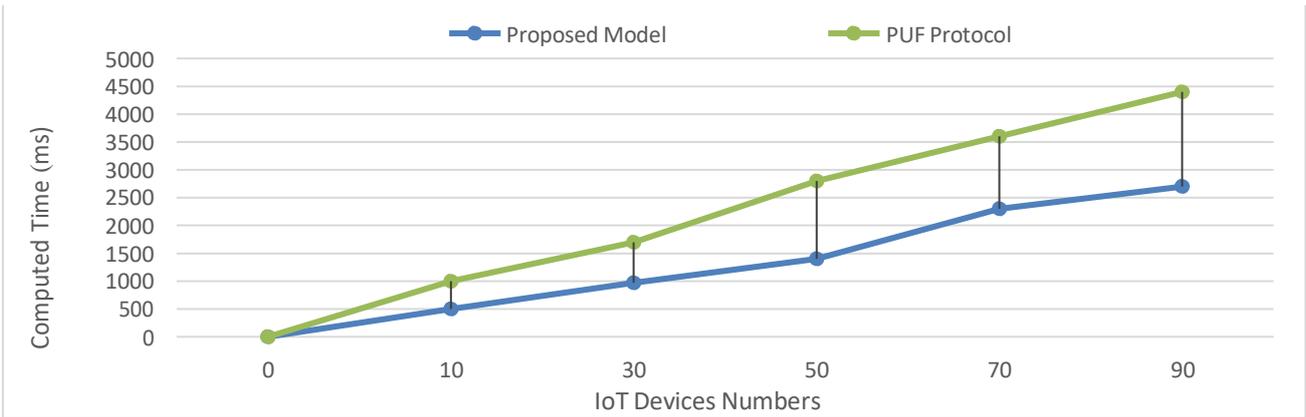

Figure 12: The computed time comparison between the proposed model and PUF protocol according to the number of IoT devices

Therefore, there will be a certain treatment methodology to the devices with low trust value were designed to be higher. If the mis-detected adversary is having a low trust lower than 0.5, then the authorization access level will be lower in accessing the services/resources in the IoT networks; in order to protect the IoT system elements and also to reduce the security threat that may be happen by the authorized devices.

The user has the ability to customize both numbers and thresholds of the authorization levels. Each customization differs from system to another based on IoMT application particular scenario. As shown in figure 13, the accuracy performance evaluation shows no false positive for IDUL.

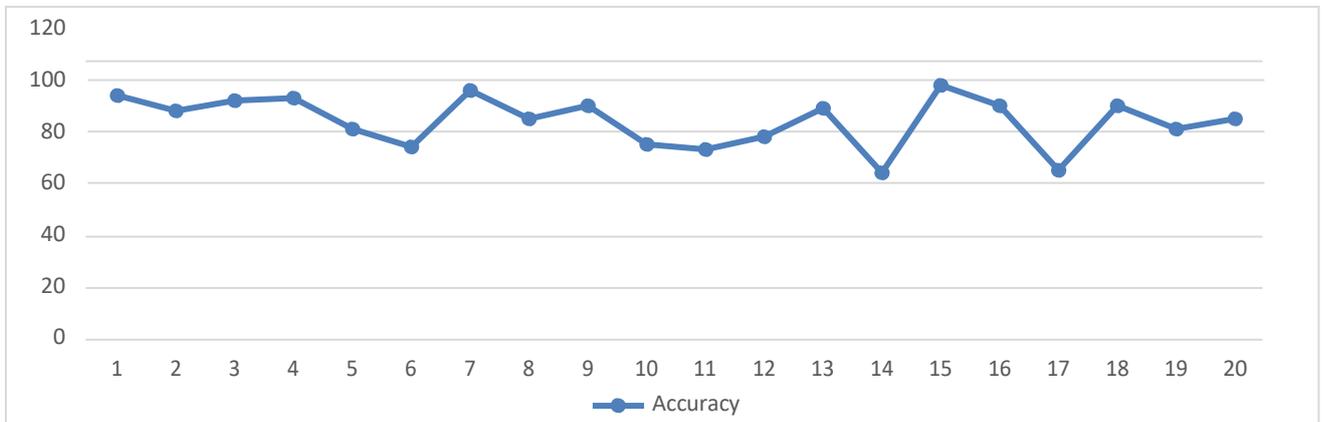

Figure 13: The accuracy performance evaluation shows no false positive for IDUL

## 7  Discussion and Analysis

Specifically, malicious attacks represent the attacking behaviors, those behaviors can be controlled by switching them on or off and the switch part will be based on applied communication nature in the IoMT systems. The creation of preliminary trust value is based on the receiver's IP address evaluation. There is a possibility for the sender or the receiver to fake the IP address.

From figure 5, it is detected that there is a noticeable fast decrement of the trust value for detected adversary by adding more attributes for the authentication. As a result, the average value was set to be with a value of 1.9; the main cause of this tendency is that the sender has a high confidence level of both the attacking behavior and the authentication attributes. Therefore, there will be a certain treatment approach to the devices with low trust value were designed to be higher.

If the un-detected adversary is having a low trust lower than 0.5, then the authorization access level will be lower in accessing the services/resources in the IoMT networks; to protect the IoMT system elements and to reduce the security threat that may be happening by the authorized devices. Consequently, the trusted third parts that are not consuming high resources capacity from the IoMT devices are not affecting the computational cost. However, when comparing the resource consumption results in the proposed techniques with the PUF protocol authentication technique that was presented in [84], it shows that an increment in computed time cost with the number of devices. It is clear that the proposed solution is having a lower computed time cost than the results using PUF protocol with different IoMT devices numbers. Finally, the result is a reflection for the fast and secure modelling for authenticating. However, the main aim was having a secure, fast and resource friendly model for authenticating in IoMT networks not measuring the cryptography security strength of the algorithm.

## 8 Conclusion

The IoT based healthcare technology is still in the promising development level, the integration of IoT technology in medical industry has significantly improved medical delivery and patients' treatment level. Securing the data-driven in medical care environment will likely lead to contribute in making proper decisions for patients' treatment delivery; to empower physicians to accurately diagnosis the patient's health status, as well as improving the emergency situations responses. Although transforming the healthcare monitoring process to be automated would improve operational productivity, there are possible risks raised during the implementation process (i.e. Insecure Data Channels, Information Leakage, Medical Devices Un-authorized access), those challenges along with restricted regulations, can affect the growth of both IoT networking and data solutions.

In this paper, we have proposed SVM authentication approach through machine learning, the initial results were presented and it demonstrates this approach feasibility for working with various types of medical IoT devices, then other possible improvements were discussed. Nowadays, it is needed to have more than 64 bits of security. Therefore, the objectives of future work are involving in extending in authenticating other domains (i.e. Physiological Biometrics and Behavioral Biometrics), it can include verification of both voice and iris.